\documentclass[]{pasj01_revised}
\usepackage{graphicx,color}
\usepackage[OT2,T1,OT1]{fontenc}
\usepackage{lmodern}
\usepackage{natbib}

\def\uas{~$\mu$as}
\def\ch3oh{CH$_3$OH}
\def\h2o{H$_2$O}
\def\arcsec{$^{\prime\prime}$}
\def\arcdeg{$^{\circ}$}
\def\kms{~km~s$^{-1}$}

\begin{document} 
%\Received{}%{yyyy/mm/dd}
%\Accepted{}%{yyyy/mm/dd}
%\Published{yyyy/mm/dd}

\title{Radio Astrometry towards the Nearby Universe with the SKA}

%%% begin:list of authors
% Do NOT capitalize all letters in "textsc".
% Alphabetical order?

\author{Hiroshi \textsc{Imai}\altaffilmark{1}%
%\thanks{Example: Present Address is xxxxxxxxxx}
}

\author{Ross A. \textsc{Burns}\altaffilmark{1}%
%\thanks{Example: Present Address is xxxxxxxxxx}
}

\author{Yoshiyuki \textsc{Yamada}\altaffilmark{2}%
%\thanks{Example: Present Address is xxxxxxxxxx}
}

\author{Naoteru \textsc{Goda}\altaffilmark{3}%
%\thanks{Example: Present Address is xxxxxxxxxx}
}
\author{Tahei \textsc{Yano}\altaffilmark{3}%
%\thanks{Example: Present Address is xxxxxxxxxx}
}

\author{Gabor \textsc{Orosz}\altaffilmark{1}%
%\thanks{Example: Present Address is xxxxxxxxxx}
}

\author{Kotaro \textsc{Niinuma}\altaffilmark{4}
%\thanks{Example: Present Address is xxxxxxxxxx}
}

\author{Kenji \textsc{Bekki}\altaffilmark{5} (SKA Japan Astrometry Science Working Group)}
%%% end:list of authors

%%%% Institution list
\altaffiltext{1}{Science and Engineering Area of the Research and Education Assembly , Kagoshima University}
\email{hiroimai@sci.kagoshima-u.ac.jp}

\altaffiltext{2}{Kyoto University}

\altaffiltext{3}{JASMINE Project Office, National Astronomical Observatory of Japan}

\altaffiltext{4}{Yamaguchi University}

\altaffiltext{5}{International Centre for Radio Astronomy Research, the University of Western Australia}

%% `\KeyWords{}' always has to be placed before `\maketitle'.
 %Do NOT move this preamble from here!

\maketitle

\begin{abstract}
This chapter summarizes radio astrometry in relation to current very long baseline interferometry (VLBI) projects and describes its perspectives with the SKA. The scientific goals of the astrometry with the SKA have been discussed in the international and Japanese communities of researchers, whose major issues are shown here. We have demonstrated some of the issues, such as censuses of possible targets and the technical feasibility of astrometry in the SKA frequency bands. The preliminary results of our case studies on SKA astrometry are also presented. In addition, possible synergy and {\it commensality} of the SKA astrometric projects with those in the optical and infrared astrometric missions, especially JASMINE (Japan Astrometry Satellite Mission for INfrared Exploration) are discussed. 
\end{abstract}

\begin{figure}
\vspace{-167mm}%%%ここの数値は体裁を変えない値を手探りで探す
\includegraphics[width=25mm]{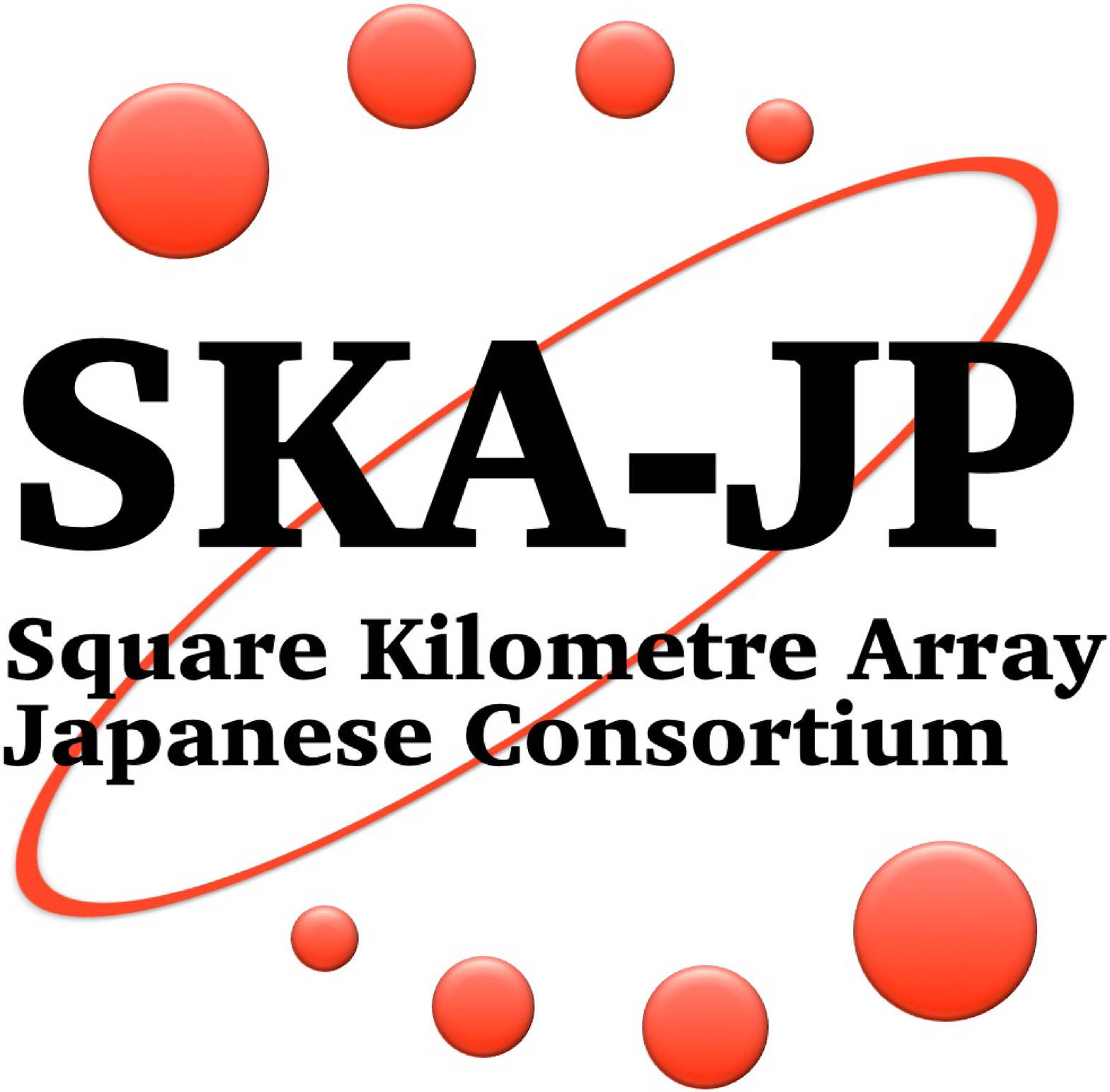}
\vspace{133mm}%%%ここの数値は体裁を変えない値を手探りで探す
\end{figure}

\section{Introduction: Radio Astrometry as Astronomical Basis and New Challenges}
\label{sec:introduction}
 
\begin{figure*}[t]
\begin{center}
\FigureFile(0.8\linewidth,100mm){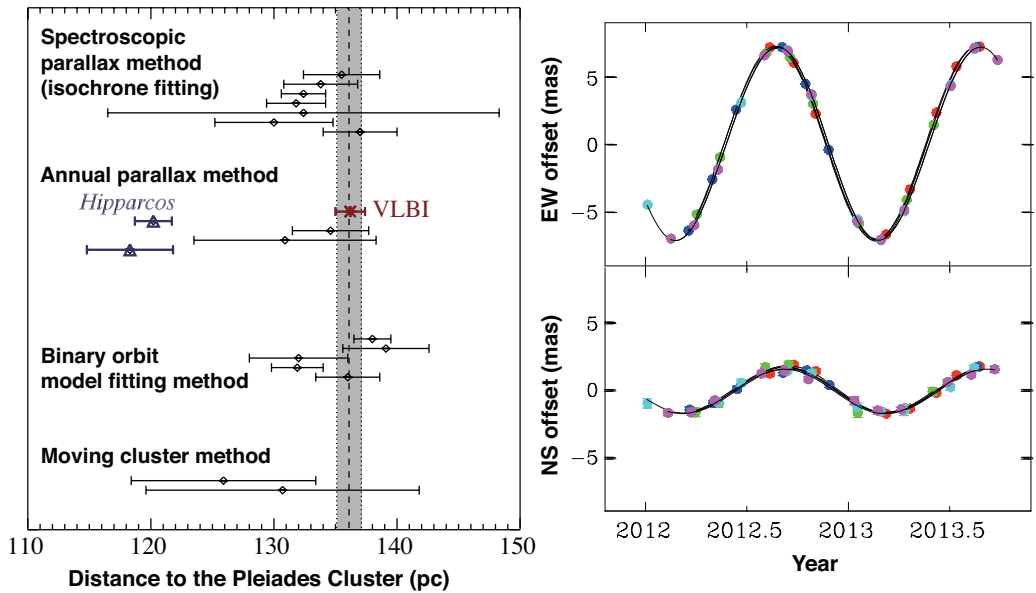}
\end{center}
\caption{Results of the distance measurement of the Pleiades Cluster summarized in \citet{2014Sci...345.1029M}. {\it Left}: Summary of the distances derived from different techniques. {\it Right}: Annual parallax modulations derived from five radio stars in the cluster, which was measured with the VLBA. The motions of the five stars are superimposed so that the derived modulations are best superimposed.}\label{fig:Pleiades-Cluster}
\end{figure*}

Astrometry serves as a backbone of astronomy and astrophysics. It has yielded trigonometric measurements of sources in the nearby universe including our Milky Way Galaxy (MWG) and the Local Group (LG) of galaxies. Annual parallax provide a crucial rung on the ``distance ladders" on which other distance estimates and standard candles are calibrated. The influences of astrometry can therefore reach cosmic scales and sometimes provides great impact on astrophysics.It has also provided precise celestial coordinate systems such as the International Celestial Reference Frame (ICRF) in radio astrometry basing on extragalactic quasars. 

\begin{figure*}[t]
\begin{center}
\FigureFile(\linewidth,100mm){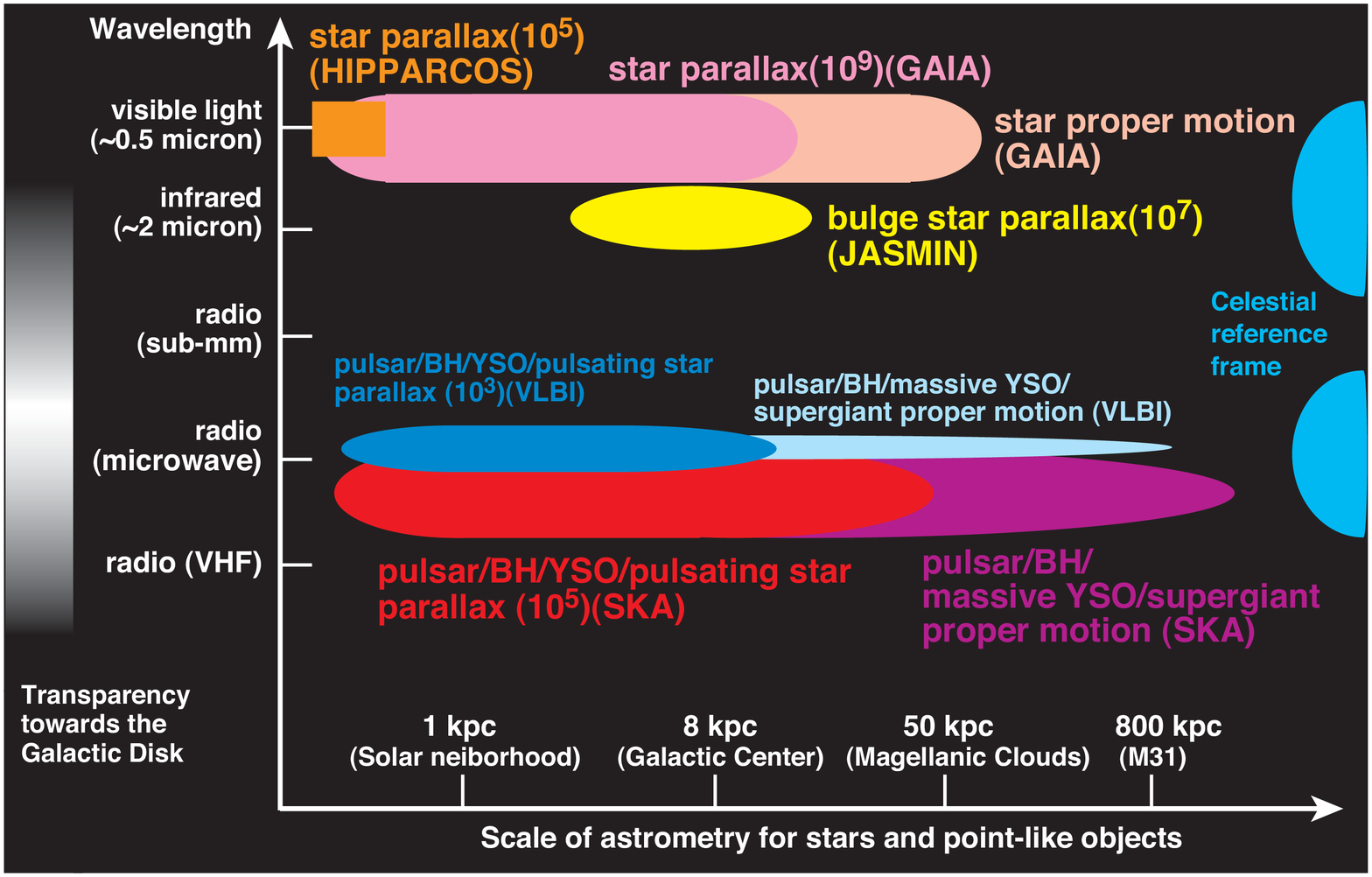}
\end{center}
\caption{Present to future view of astrometric missions. One can see optical, infrared, and radio astrometric missions, which target the different orders of magnitude of and different sorts of celestial objects. These missions shall link to each other with common targets, i.e., extragalactic quasars in the celestial reference frames and nearby bright stars emitting a wide range of electromagnetic waves such as radio stars and young and evolved stars.}\label{fig:astrometric_missions}
\end{figure*}

Fig. \ref{fig:astrometric_missions} shows the present to future view of astrometric missions. Radio astrometry is at a unique position with regards to other approaches to astrometry. The numbers of the targeted sources in the present radio astrometry missions ($N\sim 10^3$ radio sources) are much smaller than those in optical and infrared astrometry, such as the Gaia mission ($\sim 10^9$ stars)(e.g. \citealt{2012A&A...538A..78L}). However, the former targets ``exotic" sources emitting non-thermal emission such as pulsars, blackholes, and masers, complementing well to the latter. One also would remember the great impact on the distance scale controversy of the Pleiades Cluster (Fig. \ref{fig:Pleiades-Cluster}, \citealt{2014Sci...345.1029M}). The annual parallaxes measured in the cluster in radio finally gave a strong constraint on the distance scale ($D\approx$136~pc), which had an unacceptable uncertainty among different results from optical astrometry. Because radio emission is relatively transparent, even taking into account stronger interstellar scattering and poorer angular resolution in longer wavelengths, it has a unique benefit to astrometry for sources embedded in gas and dust in the Galactic plane. The radio astrometry projects in very long baseline interferometry (VLBI), using the Very Long Baseline Array (VLBA) (e.g., Bessel project, \citealt{2011AN....332..461B}), the European VLBI Network (EVN), the Long Baseline Array (LBA), and Japanese VLBI Exploration of Radio Astrometry (VERA) \citep{2003ASPC..306P..48K} have determined the locations of a few hundred maser sources in the MWG, which trace the spiral arm patterns of the  galaxy, and investigated the dynamics of the galactic disk \citep{2014ARA&A..52..339R}. 

In the era of the SKA, which will detect much fainter radio sources, we expect that one to two orders of magnitude higher number of the radio sources will be targeted in radio astrometric observations and they will cover a much larger variety of sources, including compact sources of thermal emission. In this chapter, we describe the plausible future of astrometry in the SKA era. Note that this chapter also aims to highlight uses of the SKA from a Japanese point of view, some of which focuses our interest on possible synergy with the JASMINE (Japan Astrometry Satellite Mission for INfrared Exploration) project (i.e. \citealt{2015IAUGA..2247720G}). 

\section{Technical concepts of the SKA Astrometry}
\label{sec:technical_concept}

\subsection{General technical requirement for the SKA VLBI for high precision astrometry}

The technical specifications of the SKA VLBI are summarized in several documents, e.g.,  {\it SKA PHASE 1 VLBI CLARIFICATION ECP 140008 ANALYSIS DOCUMENT} \citep{SKA-VLBI}; {\it SKA PHASE 1 SYSTEM (LEVEL 1) REQUIREMENTS SPECIFICATION} (\citealt{SKA1-requirement}; \citealt{2015aska.confE.143P}). Here we revisit those issues from a point of view of high precision radio astrometry to consider possible science goals (discussed in \ref{sec:science}) which may be approached with the SKA. 

The astrometric accuracy,  $\sigma_\theta $, which is  determined by the sensitivity (signal-to-noise ratio on the image $R_{\rm SN}$) and the longest baseline $B_{\rm max}$ is described as \citep{1993LNP...412..244M} 

\begin{equation}
\sigma_\theta \approx 0.5\frac{\theta_{\rm beam}}{R_{\rm SN}}
\approx 1000\frac{\lambda/{\rm[10~cm]}}{(B_{\rm max}/{\rm 1000~km})(R_{\rm SN}/10)} [\mu{\rm as}]. 
\label{eq:accuracy}
\end{equation}

\noindent
In SKA astrometry, $\lambda=$1--100~cm ($\nu=$ 300~MHz--30~GHz) is expected\footnote
{The higher frequency in SKA1-MID is determined to be 15~GHz, but the SKA Advanced Instrumentation Program (AIP) considers 
up to 25~GHz.}. For the state-of-the-art astrometry ($\sigma_\theta\leq $10\uas), very high sensitivity ($R_{\rm SN}>>$100) is needed. 

The position-reference sources should have very high brightness temperatures ($>>10^{6}$~K), as seen in quasars. 
As quasars emit continuum emission, their detection sensitivity is improved by using signal recorders with larger band widths.
The number density of quasars on the sky 
(section \ref{sec:reference_sources}) allows calculation of the band widths required for high precision astrometry.

One also considers the detectable flux densities and brightness temperatures independently for continuum and spectral line sources. The former sources include pulsars, the main targets in the SKA, and non-thermal sources associated with blackholes and a variety of radio stars in the MWG (section \ref{sec:spiral_arm_tomography}), super nova explosions, and active galactic nuclei in the nearby universe. The latter sources include maser sources: hydroxyl (OH); methanol (\ch3oh), water (\h2o), ammonia (NH$_3$), associated with young and evolved stars in the MWG and the LG.Regarding SKA astrometry, in addition, we can consider not only such non-thermal sources but also {\it thermal emission} sources by tracking their clumpy structures. They include radio recommbination lines and molecular lines that are relatively isolated. Although they will be detectable only with the SKA core telescopes ($B_{\rm max}\leq$160 km) and the astrometric accuracy will be limited, measurement of their secular motions and statistical-parallax distances should be scientifically attractive. This will showcase the diversity of SKA astrometry. Fig. \ref{fig:astrometric_missions} shows the numbers of target sources in radio astrometry missions, which should be more precisely evaluated in observations with the ``SKA Pathfinders" and ``Precursors". Note that the thermal sources may be brighter at shorter wavelengths, however going to shorter wavelengths contracts the field of view, reducing the number of available in-beam reference sources. Receiver sensitivity may also decrease. 

For VLBI, one has to consider two types of sensitivity. The first is a baseline sensitivity, the minimum detectable flux density above 10-$\sigma$ noise level $S_{\rm min}$, which is necessary to obtain solutions of visibility calibration. In the SKA, it is calculated between the core and a remote station as, 

\begin{eqnarray}
S_{\rm min} & = & R_{\rm SN}\frac{\sqrt{SEFD_{\rm core}SEFD_{\rm rmt}}}{\sqrt{2\Delta\nu\tau}}
\nonumber \\
 & \approx & 0.55\frac{\sqrt{[SEFD_{\rm core}/{\rm 3~Jy}][SEFD_{\rm rmt}/{\rm 100~Jy}]}}{\sqrt{[\tau/100~s]}} [{\rm mJy}], \nonumber 
\label{eq:baseline-sensitivity}
\end{eqnarray}

\noindent
where $D_{\rm core}$ and $D_{\rm rmt}$ are the effective diameters of the core and remote stations, respectively, $R_{\rm SN}=10$ is the signal-to-noise ratio, $SEFD$ the system equivalent flux density [Jy], meaning the antenna sensitivity, $\Delta\nu\approx 5\times 10^8$[Hz] the recording bandwidth (for a continuum reference), and $\tau$ the coherent integration time. 
The second is an array sensitivity, which is approximated to that yielded by baselines between the core and the remote stations.  
In the case of a maser line, it is calculated as, 

\begin{eqnarray}
S_{\rm line} & = & S_{\rm min, \Delta t}\sqrt{\frac{\Delta t}{T_{\rm total}}}\frac{1}{\sqrt{N_{\rm rmt}}} \nonumber \\
& \approx &
16\frac{\sqrt{[SEFD_{\rm core}/{\rm 3~Jy}][SEFD_{\rm rmt}/{\rm 100~Jy}]}}
{\sqrt{[\Delta\nu/{\rm 10~kHz}])[T_{\rm total}/{\rm 600~s]})[N_{\rm rmt}/{\rm 10}]}}[{\rm mJy}], \nonumber
\label{eq:array-sensitivity}
\end{eqnarray}

\noindent where $N_{\rm rmt}$ is the number of the remote stations and $T_{\rm total}$ the total integration time. 
We expect that even the SKA Phase 1 (SKA1) VLBI will be able to make astrometry for maser sources as faint as 0.1 Jy, fainter by a factor of 10 than the MWG maser sources targeted by the present astrometric VLBI projects. The sensitivity is improved by a factor of 1.7 when including baselines between $N_{\rm rmt}=$40 remote stations.  

Table \ref{tab:specification} gives the basic specification parameters relevant to SKA VLBI and astrometry. 

\begin{table}[h]
\caption{Parameters expected in astrometry with the SKA MID band. See also \citet{SKA1-BD}.}\label{tab:specification}
\begin{center}
\vspace{-3mm}
\begin{tabular}{lrr} \hline\hline
 & Band 2 & Band 5 \\ \hline 
Frequency coverage &  0.95--1.76~GHz & 4.6--13.8~GHz \\
Coherence time\footnotemark[1] & 200~s & 100~s \\
Spectral resolution $\Delta \nu$ & 3.9~kHz &  9.7~kHz \\
FoV radius\footnotemark[2] & 0.49$^{\circ}$ & 0.07$^{\circ}$ \\
$SEFD_{\rm core,SKA1}$\footnotemark[3] & 3.0 Jy & 4.0 Jy \\
$SEFD_{\rm core,SKA2}$\footnotemark[4] & 1.0 Jy & 1.3 Jy \\
$SEFD_{\rm remote,2016}$\footnotemark[5] & 100 Jy & 390 Jy \\
$SEFD_{\rm remote,SKA2}$\footnotemark[6] & 16 Jy & 21 Jy \\
$N_{\rm base, SKA2}$\footnotemark[7] & 40 & 40 \\
\hline
\end{tabular}
\end{center}

\noindent
Assumptions: \\
\footnotemark[1]Empirically selected. \\
\footnotemark[2]FoV before beam-forming. \\
\footnotemark[3]60\% of the collecting area of the core station ($B_{\rm max}\leq $100~km). \\
\footnotemark[4]30\% of the SKA2 collecting area, 10 times as large as SKA1. \\ 
\footnotemark[5]60~m (Band 2) and 30~m (Band 5) of antennas in diameter, which are commissioned in 2016, 
with an aperture efficiency of 50\% and a system temperature of 50~K. \\    
\footnotemark[6]Remote stations composed of 25 15~m antennas. \\
\footnotemark[7]To use half of SKA2 antennas to form the remotes. 
\end{table}

\begin{figure*}[t]
\begin{center}
\vspace*{-2mm}
\FigureFile(0.35\textwidth,100mm){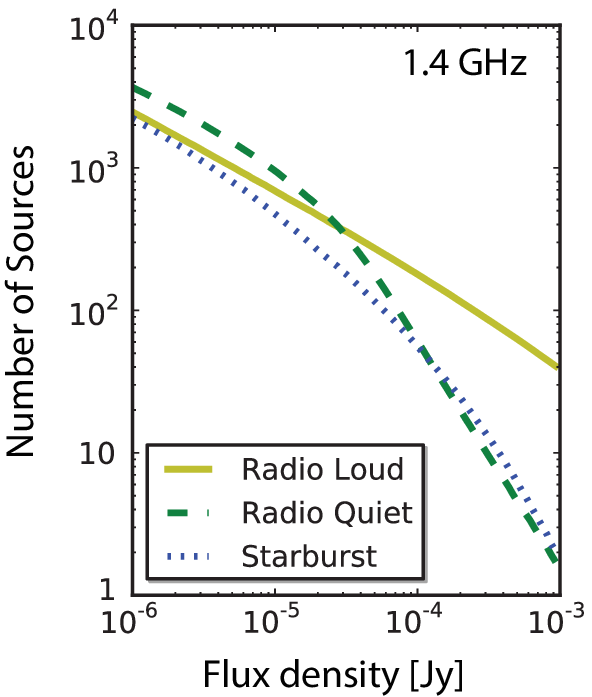}
\FigureFile(0.47\textwidth,100mm){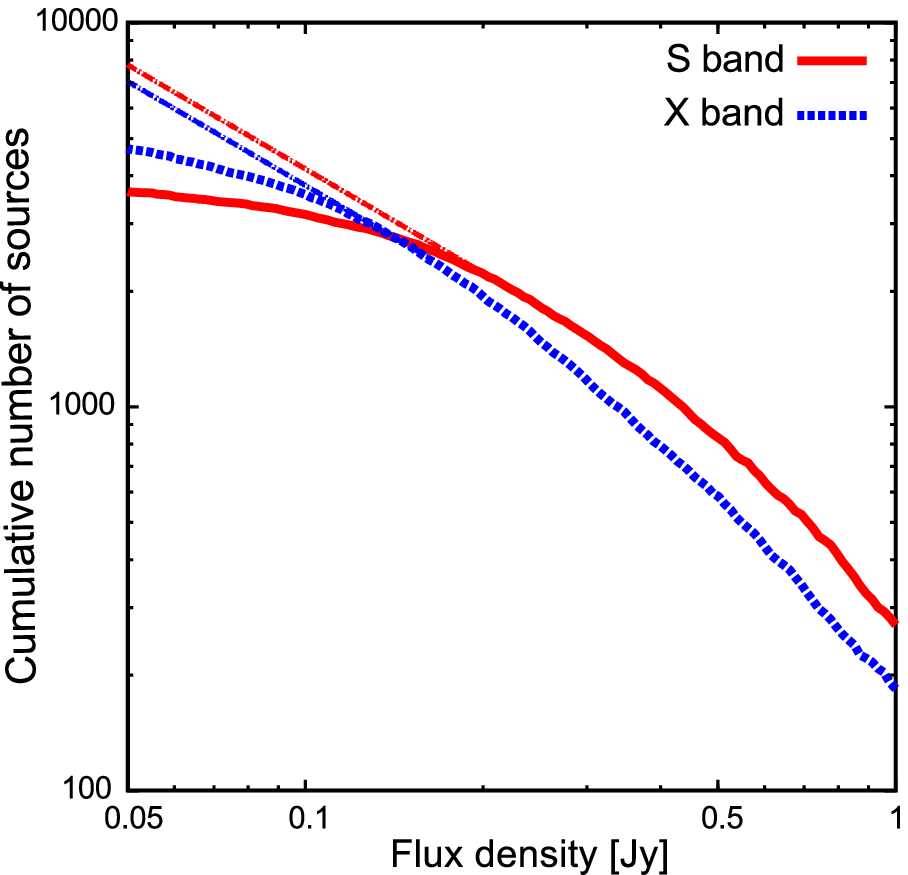}
\end{center}
\vspace{-2mm}
\caption{{\it left}: Number of possible radio sources in the SKA-MID antenna beam \citep{SKA135}. The gradient of the number count function, which includes radio loud sources such as QSOs ($N\propto S_{\nu}^{-0.9}$), can be used for estimating the number of possible reference sources with VLBI. {\it Right}: Cumulative histogram of the VLBI calibrators registered by 2015 July (rfc2015c, astrogeo.org) at $\delta \ge -30^{\circ}$. The dot-dashed lines have the same slope as on the left (solid green). One can estimate the total number of VLBI calibrators brighter than 1 mJy by extrapolating using this line ($\sim$10$^{5.4} \approx 250\,000$ sources at the S-band).}
\label{fig:SKAcalibrators}
\end{figure*}

\subsection{Special features of VLBI equipment in the SKA}

In the SKA, we consider either ``in-beam astrometry", in which target and reference sources shall be observed in a common field of view or antenna beam, or ``multi-view astrometry" \citep{2013AJ....145..147D}, in which they are observed simultaneously in multiple antenna beams. Here some expected special features of VLBI equipment in the SKA are described. 

\noindent{\bf Wide band receiving system:} The SKA Advanced Instrumentation Program (AIP) considers new designs for wide-band single pixel feeds (WBSPF) to cover as wide frequency coverage as possible in a single band. It is important to expand to lower (down to 1~GHz) as well as higher frequencies for calibrating the effects of the ionosphere in astrometry. The state-of-the-art technology covers frequency ranges of 1--20~GHz and 2--20~GHz in feeds and receivers  (e.g., \citealt{Dubrovka2010,Komiak2011,Ujihara2014}). 

\noindent{\bf Beam forming:} It is requested to form {\it at least four beams}\footnote
{the SKA1 Level 1 System Requirement revised recently mentions four VLBI beams {\it at maximum}, therefore submission of an Engineering Change Proposal (ECP) will be considered.} in each station to simultaneously observe a target and at least 
three references for calibrating the effects of the atmosphere (mainly the ionosphere and the wet component in the troposphere). 
This is {\it always} necessary, even for the in-beam astrometry, for solving the dependence of these effects on antenna elevation\footnote
{This concept was also considered in the 1990's plan of VERA(VLBI for Earth Rotation and Astrometry)}. Note that the pulsar timing array (PTA) in the SKA considers a similar beam forming concept, therefore that for VLBI should have compatibility with that for the PTA. 

\noindent{\bf Frequency standards:}
Dark fiber networks shall be deployed to the whole core station (within 100~km) and nearby remote stations to provide reference signals of the frequency standards in the core center. One must consider a cost-performance trade-off between putting the standards in the individual remote stations and deploying a fiber network over thousand kilometers in the future. 

\noindent{\bf Recorders} The signal recording rate gives a constraint on an available bandwidth. A total bandwidth of 1~GHz or larger may be available in the SKA VLBI, but the bandwidth will be determined by the sensitivity of the receivers at a lower frequency band and the specification of the optical fiber networks for the intercontinental remote stations. 

\subsection{Data calibration and astrometry}
\label{sec:in-beam-astrometry}

The present calibration schemes in VLBI are considered, including standard procedures such as amplitude calibration, fringe fitting, and self-calibration,  and advanced ones specific to high precision astrometry. The latter calibrations should take into account the ionosphere and the troposphere, whose effects depend on antenna elevation, using the multiple fields of view. Installation of GPS systems in the individual remote stations may be beneficial as seen in the VERA astrometry \citep{2008PASJ...60..951H} even if the multiple beams and in-beam astrometry will be realized. 

Moreover, further analysis of radio images is needed for high accuracy astrometry. In optical / infrared space astrometry such as JASMINE, we solve characteristics of instruments and astrometric parameters simultaneously.  For instruments, we make a precise model of a point spread function (PSF), evaluate the position of stellar center on the detector in order of 1/100 pixel accuracy, and also apply enough accurate models of deformation of focal plane, characteristics of the detector and its degradation by cosmic rays, characteristics of read/out electronics, orbit and attitude of the satellite.  By solving all these models simultaneously, we can estimate parallax and proper motion with high accuracy. In such a complicated situation, one will have to consider new techniques of true images such as the ``sparse modeling" \citep{2014PASJ...66...95H}. 

Furthermore, there are issues for both optical and radio astormetry. For estimating stellar position, we can only use the photo (brightness) center.  If there are misalignment between the center of star and the photo center, it will cause a modeling error.  We should repel unresolved binaries and stars which is affected by gravitational lensing.  If the residual of an estimate does not decrease as $\sqrt{N}$, where $N$ is number of
observations, it means there are systematic errors.  For example, a radius of a super giant becomes sometimes 1AU.  If the super giant has
star spots, it will introduce a bias in the astrometric position. 

\subsection{Feasibility of astrometry at long wavelength bands}
\label{sec:low-wavelength-astrometry}

Beside of the calibration issue described in the previous sections, one must consider the limits of SKA astrometry encountered at relatively long wavelength bands. A major limiting factor is interstellar scattering by plasma in the Galactic plane. This effect is severe towards the Galactic center \citep{2015aska.confE..36K}, and astrometry at 8~GHz or higher is suggested (e.g., \citealt{2015ApJ...798..120B}). Another major factor is the intrinsic size of the targeted radio source, especially maser sources. Empirically, OH masers at 1.6~GHz are spatially resolved out on intercontinental baselines \citep{2013ApJ...773..182I,2013ApJ...771...47I}. We expect that the trade-off between a better signal-to-noise ratio of source detection and a better angular resolution will lead to a preferable baseline length of 2000--3000 km for the OH maser sources, comparable to the whole scale of the SKA Phase 2. 

\subsection{Reference sources}
\label{sec:reference_sources}

The number of available VLBI reference sources at a given flux density level can be estimated based on the presently available radio fundamental catalog (rfc2015c, astrogeo.org), with the emphasis that these estimates are all based on the northern hemisphere coverage, which is denser due to a higher number of available radio telescopes. It is therefore crucial to mention that the on-going southern VLBI surveys have to continue in order to densify the reference frame for the SKA (e.g. \citealt{2011MNRAS.414.2528P}). Based on the right panel of Fig.~\ref{fig:SKAcalibrators}, the level at the 100\% completeness of the present calibrator catalog is estimated to be $\sim$200~mJy in S-band (2.3~GHz) and $\sim$150~mJy in X-band (8.4~GHz), where the cumulative count deviates from the log$N$--log$S$ linear dependence seen up to that point. 

However, it cannot be assumed that the faint, but compact population at the expected SKA sensitivity of $\sim$1 mJ follows the same log$N$--log$S$  relation seen for brighter AGNs. According to \cite{SKA135}, the number of faint radio-loud AGNs $N$ are proportional to $S_{\nu}^{-0.9}$, where $S_{\nu}$ is the flux density at frequency $\nu$ (the left panel of Fig.~\ref{fig:SKAcalibrators}). By using this relation, we can estimate the total number of VLBI calibrators brighter than 1~mJy by extrapolating from the presently available catalog, which gives $\sim$10$^{5.4} \approx 250\,000$ sources at S-band. This gives a mean number density of calibrators on the sky to be 6 sources per square degree. An upper estimate of all observable radio sources can also be calculated from the NRAO VLA Sky Survey (NVSS) and the VLA FIRST survey. On a $\sim$1~mJy level, they both give an unresolved source density around $\sim$50 sources per square degree. This means that the maximum number of reference sources observable with the SKA must be $\lesssim$10$^6$ sources, which also agrees with our estimate. 

\subsection{Building a radio celestial reference frame}

The accuracy of alignment between Hipparcos coordinates and ICRF at present is about 0.5~mas \citep{1998A&A...331L..33F}.  The relation between the two coordinate systems also varies in time.  Gaia in optical astormetry, and VERA and VLBA in radio astrometry will achieve 10~\uas\ accuracy.  By using common sources, we will achieve 10~\uas\ alignment accuracy between optical and radio coordinates.  The problem is that at such an accuracy, one cannot consider QSOs to be at fixed points. The photo centers of radio soruces are not always the centers of mass, and will move in time if their spatial structures change\citep{2008IAUS..248..324B}.  In addition, due to the difference in optical depth, the position of AGN jets appear to change depending on the observing frequency (core shift effect, \citealt{2011Nature...477...185H}).  The Gaia mission aims to achieve statistically-determined 0.3\uas~yr$^{-1}$ non-rotating coordinates by using 500 000 QSOs. In order to achieve a radio reference frame compatible with that of the Gaia mission, we must therefore deduce precise models of jet structures, and obtain a statistical sample of astrometric results for QSOs so that we may approach a 10 \uas-level accuracy for the constructed reference frame.

\section{Science in SKA astrometry}
\label{sec:science}

The science of SKA VLBI will have great diversity, thanks to its exceptional sensitivity and excellent angular resolution, from the compact objects to key topics in astronomy, from planetary science to cosmology, which is summarized in {\it SKA1 LEVEL 0 SCIENCE REQUIREMENTS} (SKA-SCI-LVL-001X) and {\it Advanced Astrophysics with the SKA}(AASKA14). This section focuses on scientific themes based on astrometry. We discuss the topics that have already been described in AASKA14, which was published through the international meetings on the SKA during 2014--2015. Here new topics discussed and demonstrated in Japanese astronomical community (especially those in sect. \ref{sec:spiral_arm_tomography}, \ref{sec:JASMINE}, \ref{sec:LMC-SMC}, \ref{sec:transients}) are added. 

\subsection{Pulsar astrometry}

The discussion of pulsar astrometry in detail is omitted although it is one of the key science topics in pulsar studies with the SKA, which has already been summarized \citep{2015aska.confE..36K}. Pulsar astrometry shall make measurements of trigonometric parallax distances and secular proper motions independently of pulsar timing. The astrometric approach enables us to reduce free parameters such as annual modulations and orbital parameters, which should be determined in the pulsar timing data used for detecting gravitational waves. The present pulsar astrometry has achieved 10\uas -level astrometry with the in-beam astrometry technique (e.g., \citealt{2013ApJ...770..145D}), which provides a good guide for the astrometry at  long wavelength bands, including those for OH masers discussed in this chapter.

\subsection{Mapping the Milky Way}
\label{sec:MWG}

\subsubsection{The backbone structure of the Milky Way}
\label{sec:MW-backbone}

\begin{figure}[t]
\FigureFile(\linewidth,100mm){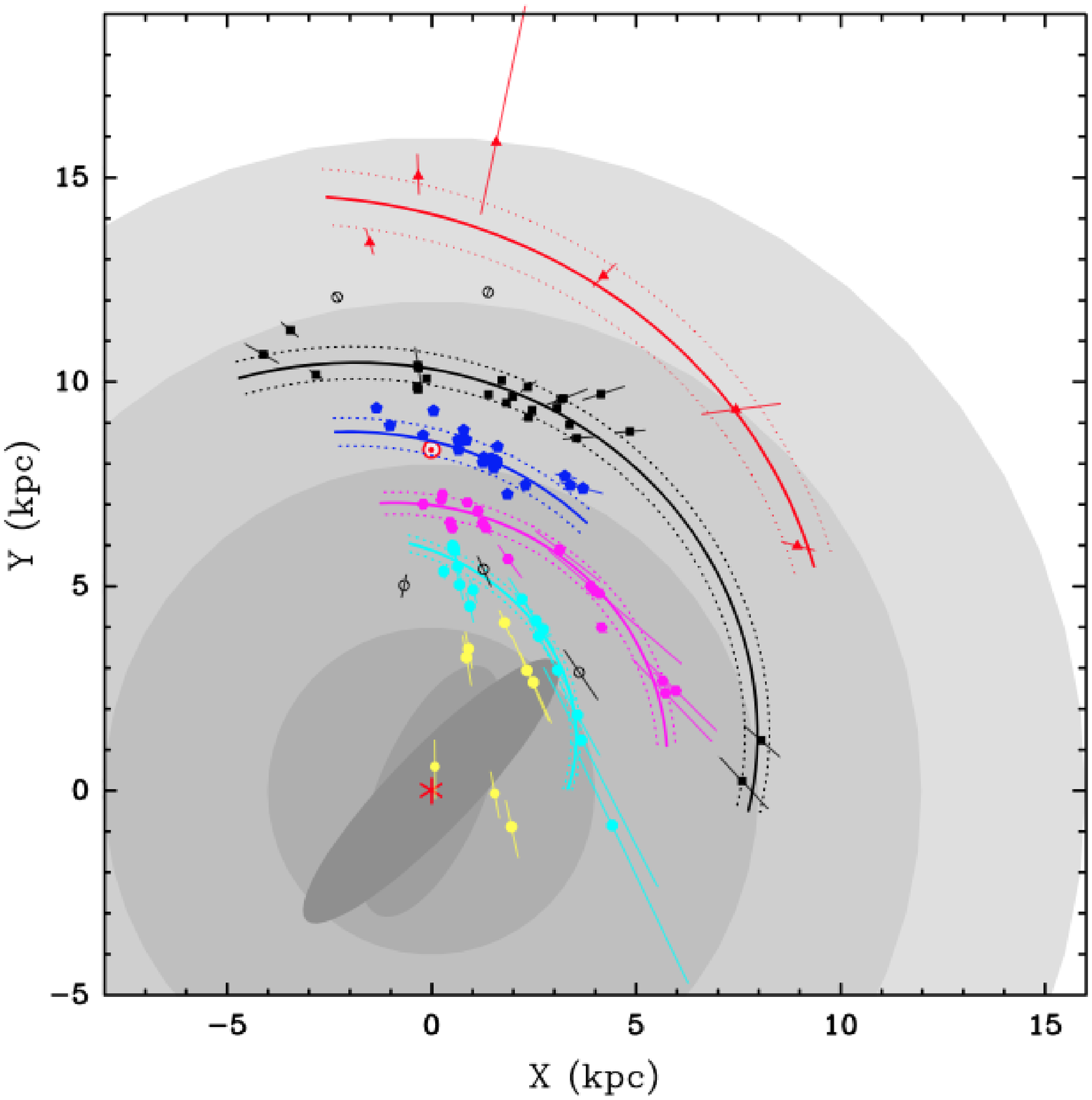}
\caption{Galactic spiral structure as evaluated using VLBI astrometric observations of star-forming regions hosting \h2o and \ch3oh maser sources 
 \citep{2014ApJ...783..130R}. Colored solid curves represent the spiral arm patterns. The gray scale schematically indicates the mean stellar density.}
\label{fig:masers_MWG}
\FigureFile(\linewidth,100mm){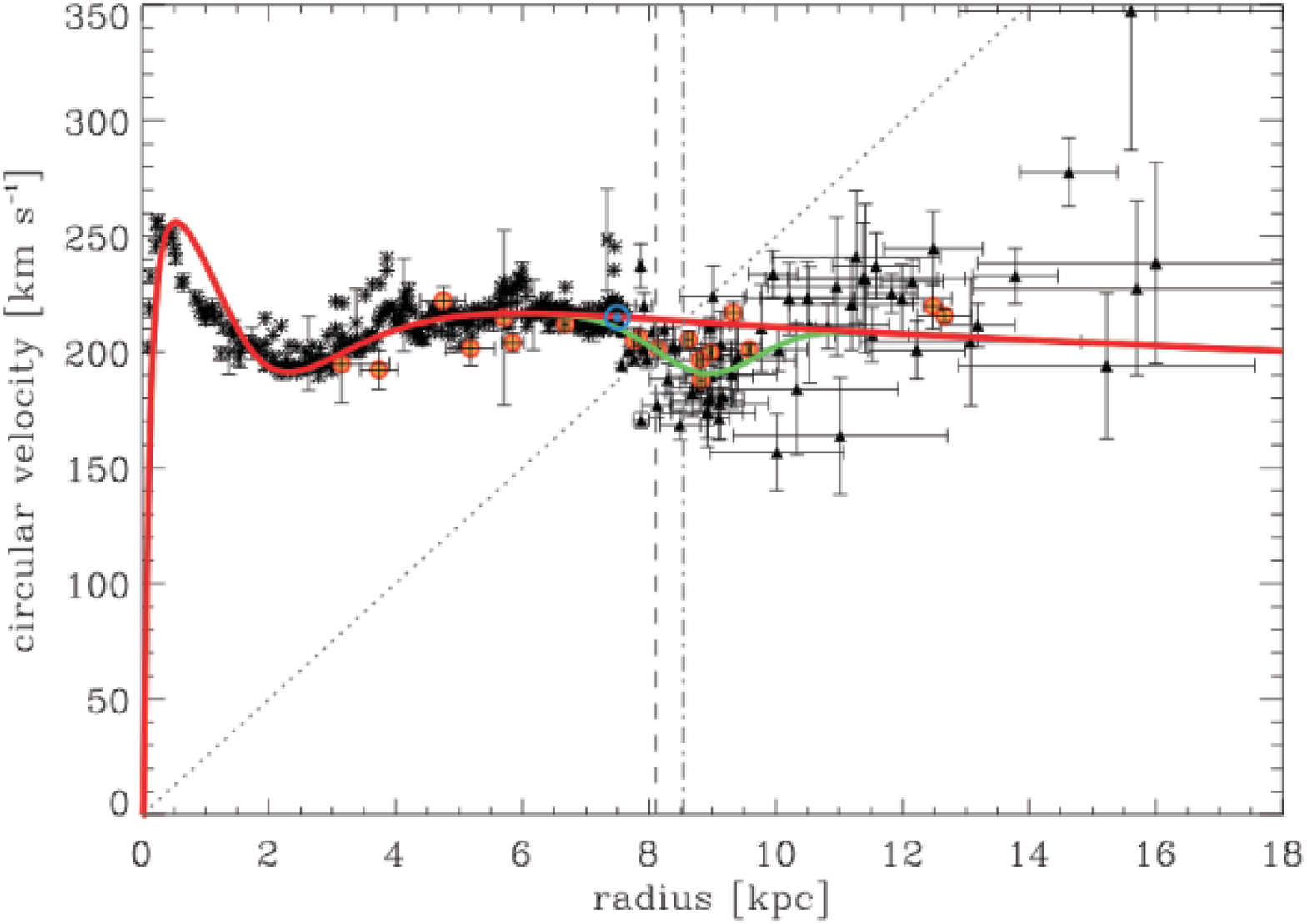}
\caption{Galactic rotation curve (a red curve) and a pattern speed of a spiral (a dotted line)\citep{2013MNRAS.435.2299B}. 
A green curve indicate a valley of the rotation curve, 
which should be formed by a possible density gap between the Perseus and the Local Arms. 
Two vertical dashed lines indicate an assumed co-rotation radii around  $R_{\rm cr} {\rm [kpc]}= R_{0}+1$.}
\label{fig:pattern_speed}
\end{figure}

In this subsection, we expand on the previous VLBI astrometry projects dealing with the MWG which were mentioned in \ref{sec:introduction}.  Accurate distance measurements for massive star forming regions (MSFRs) in the MWG have been made in these projects and enable us to visualize the spiral arms. The exclusivity of the 6.7 GHz maser line to MSFRs \citep{2013MNRAS.435..524B} makes it an ideal tracer of the spiral structures \citep{2014ApJ...783..130R}. The 22 GHz water maser sources are also commonly associated with MSFRs, though not so exclusively. Together, these maser sources cover a wide area of the MWG  (see Fig. \ref{fig:masers_MWG}). 

Currently, the number of astrometry candidate sources is typically limited by the availability of bright radio continuum sources in the nearby sky which can be used as phase reference. The increase in sensitivity with the SKA will enable in-beam astrometry for a larger number of target--reference pairs in common antenna beams and faster astrometric surveys. Eventually, one order of magnitude higher number of sources or more will become available targets for SKA astrometry. Also note that the most active VLBI arrays for astrometry at present are all located in the northern hemisphere, therefore our current best maps of the MWG are incomplete in the southern sky. SKA VLBI will be instrumental in mapping the southern sky of the MWG for solving the selection bias in determining the MWG structure. 

\subsubsection{Measuring $\Omega_{\rm p}$ and the ``spiral arm tomography"}
\label{sec:spiral_arm_tomography}
One of the plausible models to explain the spiral structures is the `density wave' model \citep{1967IAUS...31..313L}. It proposes that a spiral pattern of a galaxy is not a permanent structure populated with stars, but is actually a spiral-shaped wave where the galactic interstellar medium (ISM) is compressed. As the ISM travels through the wave, the increased gas density proliferates star formation which results in the creation of short-lived, massive stars. In this model, the ISM exhibits differential rotation in the galaxy, but is not directly coupled to the pattern rotation speed of the spiral, $\Omega_{\rm p}$, which is thought to rotate like a solid body; this property is required to solve the `winding problem' of galaxy evolution. 

The (differential) rotation curve of the MWG traced by the ISM and the solid rotation pattern speed intecept at a Galactocentric radius called the `corotation radius', $R_{\rm cr}$. As is shown in the radial velocity profile of the MWG evaluated by \citet{2013MNRAS.435.2299B} (see Fig. \ref{fig:pattern_speed}), the MWG ISM rotates more quickly than the pattern speed for $R \leq R_{\rm cr}$. Subsequently, the pattern speed is faster than the rotation speed of the ISM for $R \geq R_{\rm cr}$. Therefore, we can expect that star formation that began in the spiral density wave should be seen to evolve away from the density wave in the leading direction (in-front of the arm) for $R \leq R_{\rm cr}$ while it should trail behind the arm for $R \geq R_{\rm cr}$.

In practice, we cannot directly measure its shape of dynamics, especially $\Omega_{\rm p}$; we must infer its nature from its interaction with the ISM (i.e. \citealt{2011MSAIS..18..185G}). The most direct approach is to observe the evolutionary sequence of objects that form in the arm and move away from it.  This was tried by \citet{2005ApJ...629..825D} using a sample of 599 open clusters. By using the known age of the clusters and by reversing pure circular rotation dictated by the Galactic rotation curve, they found that most open clusters formed in arms and that the spiral pattern rotates like a solid body with $\Omega_{\rm p}= 24-26$ km s$^{-1}$ kpc$^{-1}$, and $R_{\rm cr}=(1.06 \pm 0.08) R_{0}$ kpc. However, one clear caveat is that the authors used the photometric distances to sources, which can be improved upon by measurement of trigonometric parallax distances. 

Any plan to repeat this kind of study using current VLBI networks is hindered because the accuracy of the parallactic distances obtained (typically 5 $\sim$ 20 \% ) is not accurate enough to probe for the internal distribution of sources within and around the arms. Moreover, the current sample size of trigonometric distances to MSFRs may be less than 1000 in total. On the other hand, the capabilities of the SKA (during phase: SKA1-MID), with the inclusion of existing VLBI networks (SKA-VLBI), will provide astrometric accuracy of 1\% out to sources at 2 kpc \citep{2015aska.confE.119G}, meaning an uncertainty of 20 pc. Since the width of the arms themselves are about 400 pc \citep{2014AJ....148....5V}, the SKA will allow astronomers to study the evolution of star forming regions along a line through the spiral arm. MSFRs are traced by a variety of radio sources such as CH$_3$OH, H$_2$O, and OH masers, radio recombination lines, and thermal and non-thermal continuum emission (e.g., \citealt{2013IAUS..289...36L}), which may be predominant at different evolutionary stages and become the targets of astrometry with the SKA. Although we must first precisely establish the chronology of these sources, we will be able to see the evolution through a spiral arm if the different evolutionary stages of a sample of objects are physically separated by a distance larger than the uncertainty in our parallax measurements ($\sim$20 pc). Such a situation is expected if we consider the sequential star formation mode, in which new generation star forming cores are created in the swept-up material around the H$_{\rm II}$ regions associated with older generation stars \citep{1977ApJ...214..725E}. The SKA will allow astronomers to study the evolution of star forming regions along a line through the spiral arm. This scheme of study is called ``spiral arm tomography". The different evolutionary stages of MSFRs will become spatially resolvable in the where the physical separation is larger than the uncertainty in distance estimates, i.e. $(\Theta - \Omega_{\rm p}) \times t_{\rm evol} > 20$ pc. Here $\Theta$ is the velocity of the rotating ISM and $t_{\rm evol}$ is the evolutionary timescale of our sample. In the case of MSFRs, $t_{\rm evol}$ may be defined by the difference between \emph{early} and \emph{late} massive star formation. 

Generally, the evolution from massive star forming cores to extended H$_{\rm II}$ regions takes about $t_{\rm evol}=10^{6}$ years. Therefore we can expect to see the evolution of MSFRs within an arm in the MWG where $(\Theta - \Omega_{\rm p}) \geq$  (20 pc)$/$(10$^6$~yrs) $\geq$20~km~s$^{-1}$. This value increases from zero at the corotation radius (Fig. \ref{fig:pattern_speed}). Therefore this kind of study is feasible for Galactocentric radii at least 0.8 kpc of the corotation radius, which is derived from $(\Theta - \Omega_{\rm p})/\Omega_{\rm p} \simeq (20 {\rm km~s}^{-1})/(25{\rm ~km~s}^{-1}{\rm kpc}^{-1}) = 0.8$ kpc. Here we  assume $\Omega_{\rm p} =$ 25 km s$^{-1}$ kpc$^{-1}$, taken from \citet{2011MSAIS..18..185G}. Furthermore, with the arrival of Gaia data it may be possible to substitute the \emph{late} evolutionary phase objects with MWG OB stars, thus increasing our value of $t_{\rm evol}$ and allowing us to probe $\Omega_{\rm p}$ close to the corotation radius. When a large enough sample is acquired, we can reverse this investigation to estimate $\Omega_{\rm p}$ at various points in the MWG to confirm whether it really does rotate like a solid body. The number density of the astrometric targets is also important to judge the models, such as the density wave model mentioned above and the spontaneous formation of spiral patterns (e.g., \citealt{2009ApJ...706..471B}). 

\begin{figure}[h]
\FigureFile(0.49\textwidth,100mm){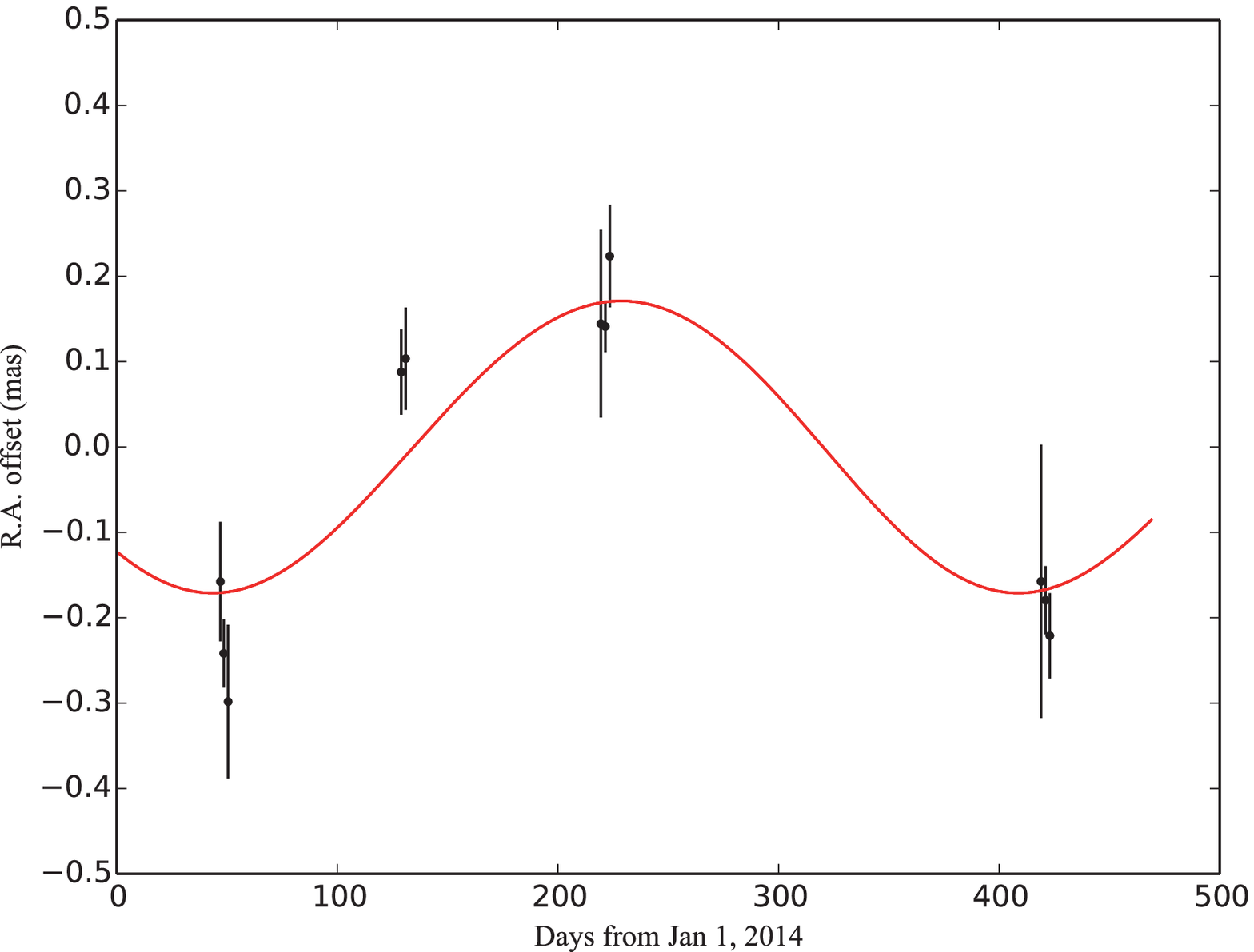}

\vspace{-2mm}
\FigureFile(0.49\textwidth,100mm){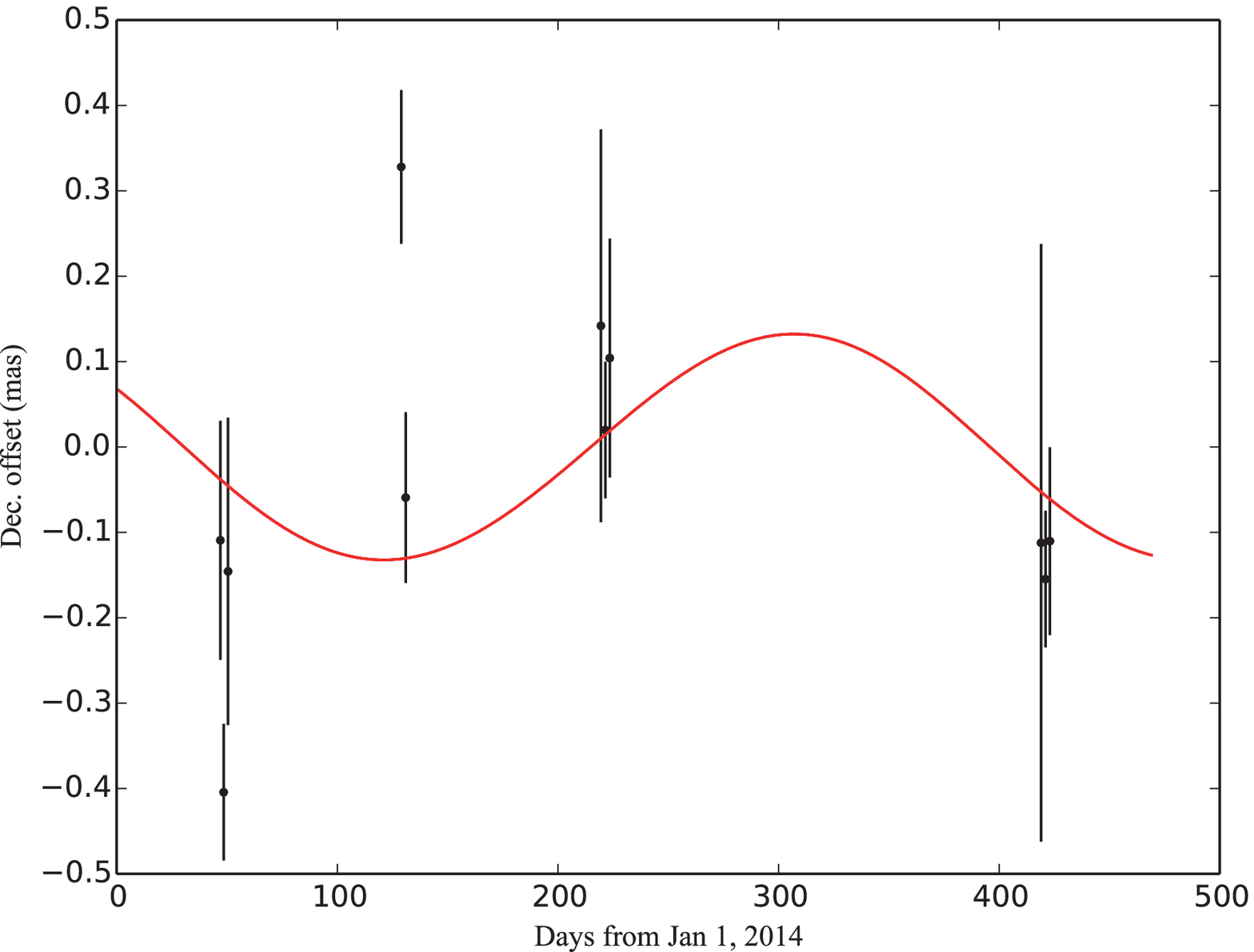}

\vspace{2mm}
\caption{A fitted parallactic motion of OH138.0+7.2 (red curve) in right ascension (top) and declination (bottom), after subtracting the linear proper motions of maser spots (Orosz et~al.\ in preparation). The maser spots in these panels are included in the maser feature with the blue-shifted peak of the spectrum and they likely correspond to the amplified stellar image of the background star \citep{2000A&A...357..945V}. Larger errors in declination are due to the characteristics of the VLBA with shorter baselines in the north--south direction. The accuracy is similar to the values from pulsar astrometry (see Fig.~1 of \citealt{2015A&A...577A.111K}.}
\label{fig:OH138parallax}
\end{figure}

\begin{figure*}[p]
\begin{center}
\FigureFile(1.0\linewidth,100mm){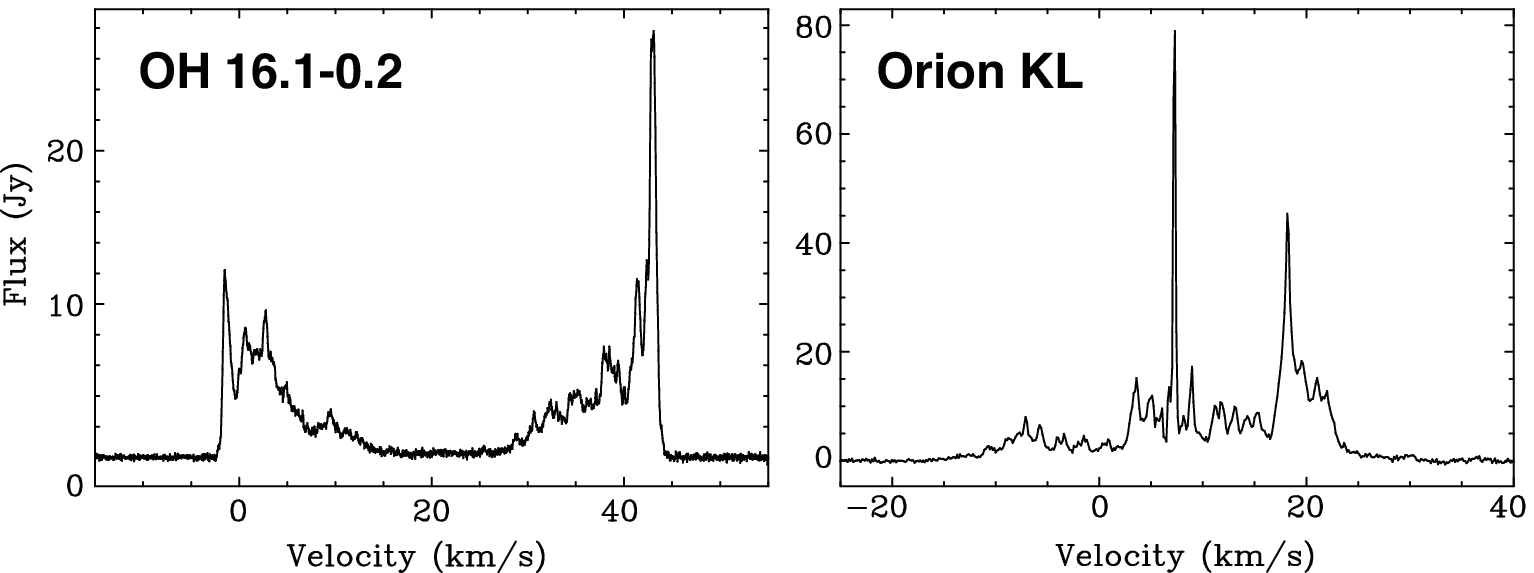}
\end{center}
\caption{Typical spectra of OH masers towards a long-period variable star (at 1612~MHz, left panel) and a MSFR (1665~MHz, right panel)  \citep{2015aska.confE.125E}.}
\label{fig:OH_spectra}

\FigureFile(1.0\linewidth,100mm){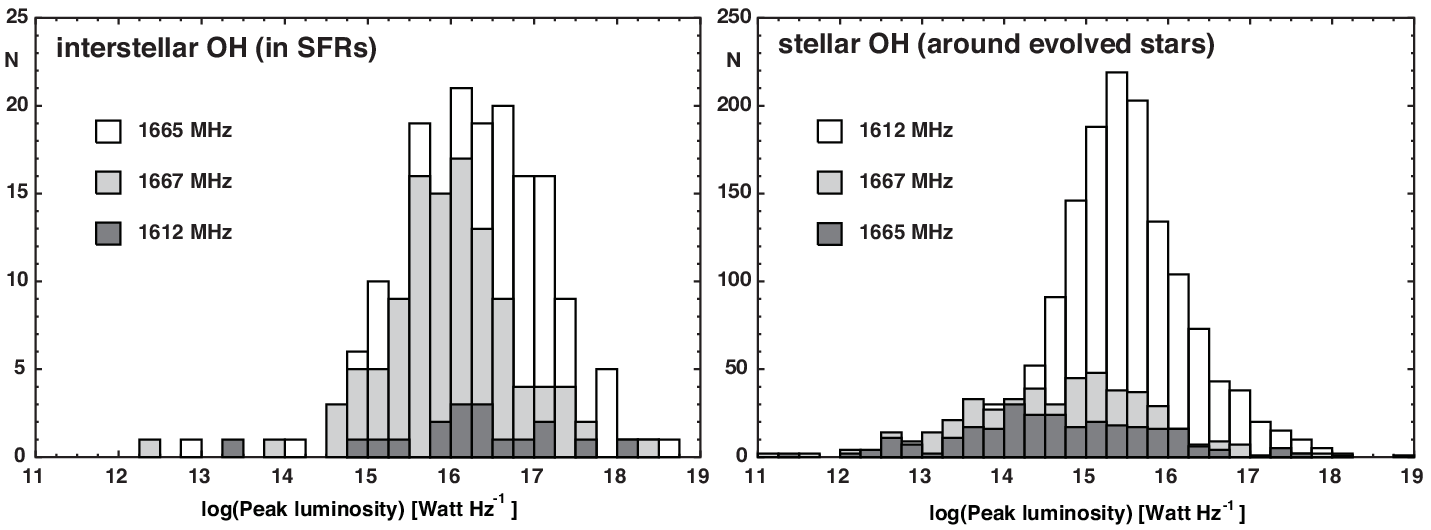}
\caption{Luminosity functions derived from known OH maser sources \citep{2015aska.confE.125E}.}\label{fig:OH_L_funcs}
%\end{figure*}

%\begin{figure*}[t]
\begin{center}
\FigureFile(1.0\linewidth,100mm){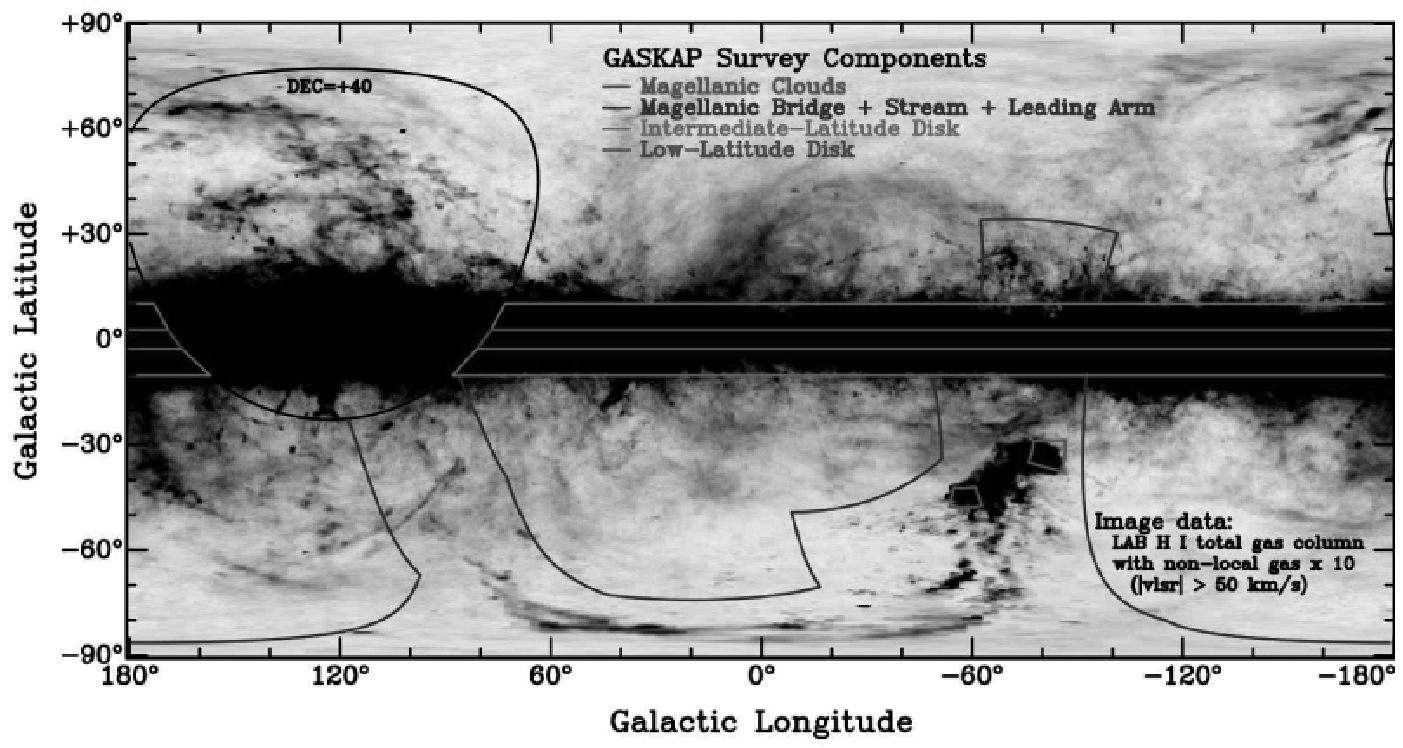}
\end{center}
\caption{Survey area of GASKAP \citep{2013PASA...30....3D}. The background grey-scale diagram shows the HI emission 
taken in an angular resolution of $\approx$600\arcsec \citep{2010A&A...521A..17K}. 
The survey area will contain the Galactic thick disk as well as a wide area of the Magellanic Cloud system.}\label{fig:GASKAP}
\end{figure*}

\subsection{Circumstellar maser sources as probes of the dynamics of the whole Milky Way}

VLBI astrometry at low frequency bands has the special difficulties described in sect. \ref{sec:in-beam-astrometry} and \ref{sec:low-wavelength-astrometry}. To correct the errors introduced by the ionosphere, we shall use an in-beam calibrator and several secondary calibrators to measure short-term dynamic variations and the two-dimensional phase gradient around the maser, respectively (the MultiView method, \citealt{2009evlb.confE..14R, 2013AJ....145..147D}). In fact, using this technique, we could determine the parallaxes and proper motions of OH masers around OH/IR stars. Fig. \ref{fig:OH138parallax} shows the preliminary results of our astrometry (Orosz et al. in prep.), which was yielded by our monitoring program of the 1612~MHz OH masers associated with the OH/IR star OH138.0$+$7.2 for a year with the VLBA. Although systematic effects still remain from the ionosphere (estimated to be $\sim$0.15~mas; cf. \citealt{2007PASJ...59..397A}), we can set a sub-mas upper limit to the parallactic motion, the smallest detected up to date using OH masers (c.f., \citealt{2000A&A...357..945V, 2003A&A...407..213V, 2007A&A...472..547V}). 1612~MHz OH masers as seen toward OH138.0$+$7.2 have a double-peaked spectrum (left panel of Fig. \ref{fig:OH_spectra}). The velocity separation between the two peaks equals to twice as large as the expansion velocity of the circumstellar envelope. For some of these sources, flux density variations of these peaks have  been precisely determined, enabling determination of the source distances using the phase-lag technique \citep{2015A&A...582A..68E}, therefore, this technique can be tested by independently derived parallax distances. 

Besides the technical considerations, the distance measurement of OH/IR stars is interesting in two major scientific issues. The first is the determination of the period--luminosity ($P-L$) relation for long-period variables (LPVs) in the MWG. So far the handful of parallax measurement all targeted short-period Mira and semi-regular variables (P$<$500 days, \citealt{2014PASJ...66..101N} and references therein), while the present experiment would be the first to determine the parallaxes of OH/IR stars with longer pulsation periods ($P=$1410 days for OH138.0$+$7.2). This enables a comparative study of various types of asymptotic giant branch (AGB) stars including the LPVs and post-AGB stars. The second is the comprehensive study on the dynamics of the whole MWG. While CH$_3$OH and bright H$_2$O masers are associated with MSFRs as mentioned in Sect. \ref{sec:MWG}, OH masers, especially the 1612~MHz OH masers, are associated with evolved stars that are widely distributed in the MWG including its bulge and halo. They are contrasted with the main line OH masers at 1665 and 1667~MHz associated with MSFRs, whose spectral profiles are different from those of the 1612~MHz line (right panel of Fig. \ref{fig:OH_spectra}). Because the luminosity functions have been well established (Fig. \ref{fig:OH_L_funcs}, \citealt{2015aska.confE.125E}), the number of detectable OH maser sources can be estimated for consideration of the SKA astrometry. 

There are $\approx$5 000 circumstellar OH maser sources \citep{2015A&A...582A..68E}, and this number will increase in the near future. In fact, the unbiased, wide area survey with ASKAP, GASKAP (Galactic ASKAP Spectral LIne Survey) program (see Fig. \ref{fig:GASKAP}, \citealt{2013PASA...30....3D}) will start in 2016, and is expected to detect more than 10 000 OH maser sources along the Galactic plane. In fact, the results of the pilot survey of OH emission, SPLASH (Southern Parkes Large Area Survey for Hydroxyl, \citealt{2014MNRAS.439.1596D}) suggests that there are $\approx$5 000 1612~MHz OH maser sources that are detectable with the SPLASH sensitivity (5-$\sigma\approx$400~mJy) along the Galactic plane ($|b|\leq 2$\arcdeg)(Shinano \etal\ in prep.). 

%________________________________________________________________

\begin{figure*}[t]
\begin{minipage}{0.65\textwidth}
\FigureFile(0.95\linewidth,55mm){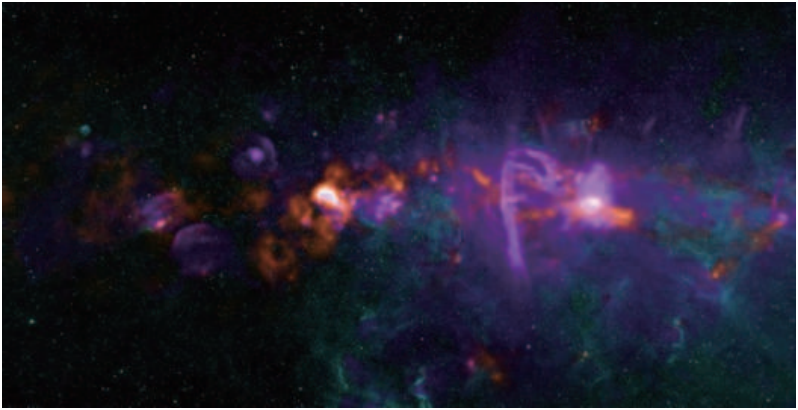}
\end{minipage}
%\hspace{-0.2\linewidth}
\begin{minipage}{0.35\textwidth}
\caption{Radio emission in the Central Molecular Zone of the MWG (@NRAO).}
\label{fig:CMZ}
\end{minipage}
\end{figure*}

\subsection{The Galactic center and the bulge: synergy of radio and near-infrared astrometry}
\label{sec:JASMINE}

  A super massive black hole (SMBH) resides in the central region of the Galactic bulge. SMBHs were first recognized as an origin of the energy at the galactic nuclei around 1960s. Subsequently it has been understood after the continuous observation that SMBHs exist universally within large galaxies. Several decades after SMBHs have been discovered, it has yet to be established as to how the SMBHs form. This makes it one of the most important problems of present-day astronomy. There are considered to be two possible formation mechanisms: gas accretion onto a seed black hole or a merger of black holes. However it is still unclear which one is the plausible explanation. From a viewpoint of the standard scenario for galaxy formation, it is natural to regard a SMBH as an end result of a BH merger. 
  
  In addition, to understand the formation and growth process of a SMBH, it is crucial to unravel the physical processes of transferring gas, stars, and intermediate-mass BHs from a bulge to a central part of a galaxy though losing their angular momenta and energy.  Since this process is essentially determined by the gravitational potential of the bulge, we need to place a high priority to the assessment of the gravitational potential of the central region of the MWG. In order to unravel these problems, it is important to search for observational relics of a sequential merger of multiple black holes to form the SMBH at the Galactic center (GC). This clarification is expected to contribute a better understanding of the co-evolution of the SMBHs and bulges in external galaxies.

\begin{figure}[h]
\FigureFile(1.0\linewidth,100mm){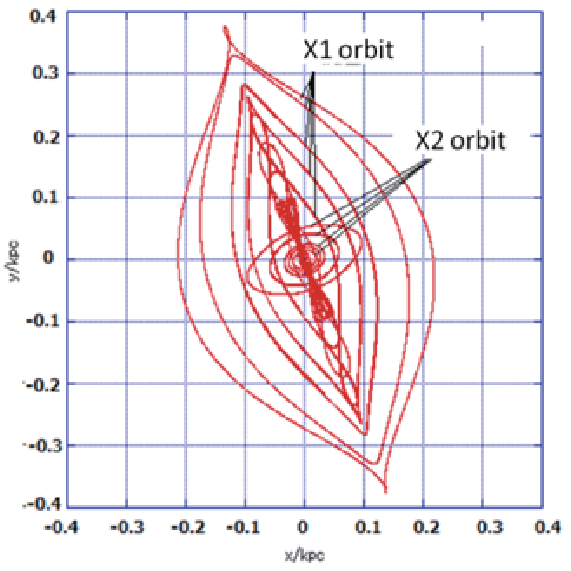}
\caption{Stellar orbits around the Galactic center. Only closed orbits (X1, X2, and inner orbits) are displayed.}
\label{fig:orbits_GC}

\FigureFile(1.0\linewidth,100mm){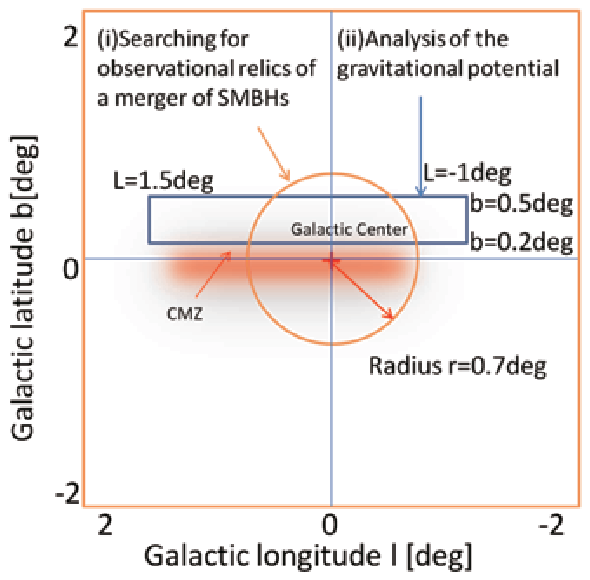}
\caption{Observing region of the Small-JASMINE.}\label{fig:Small_JASMINE}
\end{figure}

  Furthermore, containing models of the gravitational potential in the MWG nuclear bulge region (within 200--300~pc away from the MWG center), which includes the Central Molecular Zone (CMZ) (Fig. \ref{fig:CMZ}) is also important.  The gravitational potential plays an important role in determining types of stellar orbits (Fig. \ref{fig:Small_JASMINE}) which are related to the physical processes of SMHB growth mentioned above. The reorganization of the gravitational potential will make it possible to predict whether gas and/or stars are transferred to the central part and the GC will be active in the future.

In the JASMINE, an infrared astrometric satellite project designed mainly at the National Astronomical Observatory of Japan, Small-JASMINE (SJ, planned to launch in 2020's) will measure the positions and motions of stars on the surface of the celestial sphere, which are located around the GC (the nuclear bulge region). SJ will survey, at near infrared wave-lengths, the region which consists of two areas (Fig. \ref{fig:orbits_GC}): one of the areas covers a circular region with a radius of 0.7\arcdeg\ centering around the GC, corresponding to a radius of 100~pc at the distance of 8~kpc from us. The other covers a rectangular area with the Galactic longitude  $-1$\arcdeg$\leq l\leq$1.5\arcdeg, and the Galactic latitude 0.2\arcdeg$\leq b\leq $0.5\arcdeg, including the CMZ. The number of stars measured by SJ with annual parallax errors of less than 20~$\mu$as and proper motion errors of less than 50~$\mu$as~yr$^{-1}$, which are located within the observing region, is about the order of 10$^4$.

 SJ will clarify the formation process of the SMBH at the GC. In particular, SJ will determine whether the SMBH was formed by a sequential merging of multiple black holes. The clarification of this formation process of the SMBH will contribute to a better understanding of the merging process of satellite galaxies into the Galaxy, which is suggested by the standard galaxy formation scenario. A numerical simulation \citep{2014MNRAS.440..652T}  suggests that if the SMBH was formed by the merging process, then the dynamical friction caused by the multiple black holes have influenced the spatial and velocity distribution, that is, the phase space distribution, of stars located in the region within the circle with a radius of 100~pc centering around the SMBH at the GC.  The phase space distribution measured by SJ will make it possible to determine the occurrences of the merging process. SJ will determine whether the SMBH at the GC was formed through merging process of multiple black holes with 99.7\% confidence level.
 
  Furthermore, SJ will constrain models of the gravitational potential in the Galactic nuclear bulge to predict the activity of the GC in the future.  In particular, SJ will determine whether the pattern speed of the gravitational potential has a value larger than 155~km~s$^{-1}$kpc$^{-1}$ with 99.7\% confidence level. It should be remarked that this pattern speed corresponds to the critical pattern speed above which gas has been transferring effectively from the Galactic bulge to the central part of the MWG though losing their angular momentums and the GC will be active in the future. 

  There are common target sources, which are mainly circumstellar OH maser sources, between the JASMINE and the SKA. Their astrometric results will be cross-checked. For stars in the innermost part of the nuclear disk, where it is difficult for JASMINE to access due to severe interstellar extinction, the SKA astrometry will compensate the sample of those stars. 

\subsection{The dynamics of the Local Group of galaxies}

The MWG is a major member of the Local Group (LG) of galaxies, together with the Andromeda galaxy (M31). The LG also contains M33 and smaller satellite galaxies such as the Large and Small Magellanic Clouds (LMC and SMC, respectively) highlighted later. Because the LG has gravitationally formed against the expansion of the universe, the density parameter of the total energy at present, $\Omega_{\rm 0}$ is expected to determine the transverse speed of M31 and the MWG \citep{1990ApJ...362....1P,1994ApJ...429...43P}. The density profiles of the dark halos of M31 and the MWG should affect the evolutions of the satellites orbiting in these halos \citep{2001ApJ...559..754M}.  Thus the relative 3D motions between M31 and the MWG should be a key parameter to elucidate the dynamics of the LG through determining the cosmological parameter and the dark halo profiles. 

\subsubsection{Proper motion of M31}
\label{sec:M31}

The measurement of the transverse speed (proper motion, PM) of M31 is much more difficult than that of the relative line-of-sight velocity of M31 with respect to the MWG in practice. The M31's PM has been measured by using the motions of stars in 5--7 years in the outer disk, the stream, and the halo with respect to the background distant galaxies on the {\it HST} images \citep{2012ApJ...753....7S,2012ApJ...753....9V,2012ApJ...753....8V}, yielding a proper motion $(\mu_{\rm R.A.}, \mu_{\rm decl.})=42.2\pm12.3,\; 30.9\pm11.7)$[\uas~yr$^{-1}$]. This corresponds to a transverse speed of M31 of $V_{\rm tan}\approx$17\kms (after subtracting the rotation and the peculiar motion of the Sun), or  $V_{\rm tan}\leq$34\kms\ in 1-$\sigma$ uncertainty level.  These results suggest that M31 will collide with the MWG almost straightforwardly in 4 billion years. However, possible systematic errors in the PMs due to the internal motions stars in M31 should be further removed by monitoring a larger number of fields of view. 

Independently, the PMs of M31 and M33 have been measured using \h2o and CH$_3$OH masers in the galaxies. \citet{2005Sci...307.1440B} detected a proper motion of M33 to be $\mu\sim$30\uas~yr$^{-1}$. The measurement of M31's PM is ongoing after the discovery of \h2o and CH$_3$OH masers in the galaxy \citep{2010ApJ...724L.158S,2011ApJ...732L...2D}. Because the number of these masers are still small ($<10$), we will have to wait for the SKA-VLBI with higher sensitivity. M31 and M33 are still observable from the SKA cores in South Africa. Even if the SKA has to spend tens of hours per epoch per field, therefore, such extragalactic PM measurement should be conducted as one of Key Science Programs in the SKA. 

\subsubsection{Large and Small Magellanic Clouds and their surroundings}
\label{sec:LMC-SMC}

\begin{figure}[t]
\begin{center}
\FigureFile(1.0\linewidth,100mm){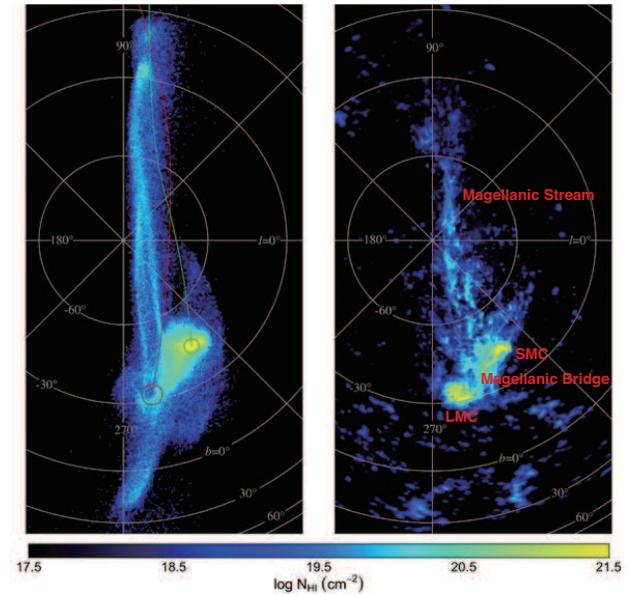}
\end{center}
\caption{Simulated distribution of gas in the Magellanic Stream (left, \citealt{2012ApJ...750...36D}) and the corresponding observational one (right, traced by H{\rm I} emission, \citealt{2003ApJ...586..170P}). These observational and theoretical results show the column density of neutral hydrogen around the Stream. The orbits of the MCs in the simulation are consistent with the orbital velocity of $\sim 300$ km s$^{-1}$ (`bound orbit') for the LMC. The fine and fainter-extended structures in the ISM distribution such as spurs are of course the targets of the SKA.\label{fig:Magellanic_Stream}}
\end{figure}

The LMC and SMC are a pair of gas-rich dwarf galaxies interacting  with each other and with the MWG.
The two Clouds (MCs) have  long  served as the best astrophysical laboratory for studies of their unique chemical evolution and star formation histories, evolution processes of interstellar medium, and the global dynamical evolution of gas-rich irregular galaxies in general, as seen in Fig. \ref{fig:Magellanic_Stream}. One of the most important parameters for the evolution of MCs is their three-dimensional (3D) motions around the Galaxy, because their interaction histories with the MWG and thus their evolution processes are determined in this way. Although many observational studies of proper motions (PMs) for the MCs have been done since 1970s, the 3D motions have not been well constrained. Recent PM studies of the LMC by HST (e.g., \citealt{2006ApJ...638..772K}) have revealed that the LMC is currently moving around 380\kms, which is significantly faster than previously thought ($\sim$300\kms). Other ground-based PM studies of the LMC have shown that the LMC's space velocities around 300\kms, which disagrees with the HST results. So, which observational result is close to the true value of the LMC's velocity?

In order to address this important question, we need to know the PMs of the LMC and the SMC much more precisely. Such precise PM measurements
could not be done in any previous optical observations because of the relatively large error bars via PM measurement ($\sim$40\kms), but, it can be done with the error of less than 10\kms\ by the SKA PM measurement. Using the SKA, astronomers will be able to measure the PMs of maser source within the LMC and the SMC so that they can estimate the 3D motions of the MCs very precisely.  This SKA PM measurement using OH masers has a number of great advantages over previous optical PM estimation. The most important factor is that a much larger number of masers sources ($\geq 200$) are available to the SKA, which enables astronomers to distinguish the secular PMs and the 3D internal kinematics of the MCs. It also should be noted that these masers are located close to the galactic disks and well follow the galactic rotation. This has a great advantage over the previous HST PM measurement, in which only 17 data points were used to derive the PM of the LMC (e.g., \citealt{2013ApJ...764..161K}).  

Thanks to these advantages of the SKA PM measurements, astronomers can much better understand the past dynamical histories of the MCs around
the Galaxy as well as the impact of the SMC-LMC-MWG interaction on the evolution of the MWG.  

\subsection{Science using the radio celestial reference frame}
\label{sec:reference_frame_science}

Gaia and the SKA will provide precise optical and radio celestial reference frames (CRF), respectively, both of which are based on thousands of extragalactic quasars, at 10\uas -level accuracy. (e.g., \citealt{2003ASPC...298...25M}). Gaia performs large-angle (up to 90\arcdeg) astrometry in order to globally determine astrometric parameters of individual ($\sim 10^9$) stars, while the SKA will do both large-angle and narrow-field astrometry using multiple beams and wide fields of view, respectively. Note that the former astrometry is sensitive to the global distortion of the CRF, such as the ``Galactic aberration" \citep{2011A&A...529A..91T}, and the latter one is sensitive to local effects such as binary and star cluster systems, ``micro-lensing" events (e.g.,  \citealt{1997AJ....114.1508H}), and interstellar scintillation (e.g. \citealt{2003ASPC..306..383J}). Although the accuracy of the CRF will be limited by those effects, reversely they will enable new space-time explorations of the universe for study on an unprecedented variety of the celestial objects.  

\subsubsection{Galactic aberration}

Independently of the astrometry for the MWG maser sources (Sect. \ref{sec:MWG}), direct measurement of the centripetal acceleration of the Sun toward the GC will determine the rotation speed of the Sun in the MWG. This acceleration should cause secular aberration drifts of extragalactic quasars, in which their acceleration vectors converge into dipoles on the sky, one of which should be towards the GC. The amplitude of the apparent proper motions of quasars due to the aberration is $\Delta \mu = V_c^2/R_0 c\approx 5$\uas~yr$^{-1}$, where $V_c$ is the rotation speed of the Sun in the MWG, $R_0$ the galactocentric distance to the Sun. Note that $\Delta \mu$ can be derived independently of the peculiar (non-circular) motion of the Sun.

\citet{2011A&A...529A..91T} obtained $\Delta \mu = 6.4 \pm 1.5$ $\mu$as~yr$^{-1}$ by using the data of 555 quasars taken during 1990--2010, with the distribution pattern of the proper motions that converges roughly into the GC. The SKA astrometry will yield a much larger size of sample and a time baselines longer than 30 years, leading to more precisely derive the Galactic abberration. 

\subsubsection{Astrometric micro-lensing events}

Optical photometric observations have found {\it photometric} micro-lensing events, in which a background star is temporarily brightened by the foreground object by a gravitational lensing effect. These events are useful for elucidating the characteristics of the foreground sources such as their masses, distances, proper motions, and binary motions affected by the existence of planets. In the case of low mass foregrounds, however, the photometric micro-lensing events are rare (40 MACHOs, \citealt{Alcock97}). 

In an {\it astrometric} micro-lensing event, on the other hand, a temporary distortion of a proper motion of the background source is found. Because a cross section of the astrometric micro-lensing is much larger than that of the photometric one (1\arcsec\ for $\sim$7\uas\ deviation, e.g., \citealt{Yano12}), the former events will be much more frequently detected. Towards the present ICRF sources behind the Galactic bulge, about one astrometric micro-lensing event of 10\uas\ per year is expected, while those of 1\uas\ will occur much more frequently, giving limitation to the accuracy of the ICRF \citep{1997AJ....114.1508H}. One can expect more astrometric micro-lensing events towards a large variety of the astrometric targets, not only MACHOs but also extra-planets, towards the Galactic bulge and the Magellanic Clouds.

\subsection{Identification of un-identified radio transients by accurate position measurement}
\label{sec:transients}

Gamma-ray bursts (GRBs), which are the most extreme high-energy transient phenomena in the universe, were discovered by continuous all sky monitor by space telescopes. Additionally, immediate follow-up observations by X-ray and optical telescopes clarified that they are associated with  distant galaxies. It means that they are cosmological phenomena (e.g., \citealt{2006Natur.440..184K}). Also conducting multi-wavelength observations and theoretical study on the afterglow of GRBs resulted in significant progress of study on the progenitors and the emission mechanism of GRBs.

At radio band, there are also several interesting transient phenomena (radio transients) such as the supernova explosions and radio bursts as seen in Cyg-X3 (e.g., \citealt{1972Natur.239..440G}). However most of radio transients have been studied by focusing on the targets, e.g., dense monitoring of several sources and the Galactic center region, or triggered by alert signals from observations at other wavelengths, because it is quite difficult to conduct wide field monitors simultaneously with existing radio telescopes. 

Time series analysis without images at radio bands (especially low frequency) has succeeded to detect transient signals, which were not well known by previous radio observations. The most characteristic property of such radio transients is their various time scales on each event. It is well known that there are two-types of lifetimes on such transient signals. One is shorter duration ($\tau<\sim$sec), and another one is longer duration ($\tau>\sim$min). As the case of short duration radio transients, Rotating RAdio Transients (RRATs: \citealt{2006Natur.439..817M}) and Fast Radio Bursts (FRBs: e.g., \citealt{2007Sci...318..777L}), which were discovered by the pulsar survey project mainly led by Parkes radio observatory, are well known. The origin of RRATs is thought to be from neutron stars in the MWG, and it shows recurrent and non-periodic strong radio pulse. On the FRBs, they are thought to have an extragalactic origin because of their large dispersion measure at high Galactic latitude. On the other hand, several types of long duration radio transients have also been discovered by several radio monitoring programs in the GC region (e.g., \citealt{2005Natur.434...50H}) and at high Galactic latitude (e.g., \citealt{2012AJ....143...96J}).

Since the origin of these radio transients except RRATs are still unknown, the origin and the emission mechanism are strenuously investigated by both observation and theory. Especially FRBs discovered by time-series analysis were expected to be cosmological phenomena because of its large dispersion measure at high galactic latitude. Since these radio transients are often found in archival data, and since they suffer from large positional uncertainties ($0.05-0.2^{\circ}$) on account of being detected with single dish telescopes, it is often difficult to identify the progenitor (e.g., host galaxies).

Therefore, in order to identify the origin and to understand emission mechanism of unidentified radio transients, in addition to the collaboration with multi-wavelength, radio telescopes which have wide fields of view, high angular resolution, and high time resolution, are desired. For example, to specify the host galaxy in which the radio transients occurred at 3 Gpc, the angular resolution of 2\arcsec\ is required. To identify the detailed location in the host galaxy, very high angular resolution less than 0.1\arcsec\ is essential.

The specification of SKA mentioned in Sect. \ref{sec:technical_concept}, will make it possible to specify the location of bright objects with an accuracy of $\sim10$ mas even if it is transient phenomenon. SKA will ensure not only higher accuracy on positional measurement to identify the counterpart but also wide field of view to search for them effectively. It is expected that the groundbreaking observational ability of SKA will give us unexpected knowledge on the origin of radio transients. 

\section{Summary}

The SKA will become an ultimate tool of time-space measurement in the universe, which enables us to understand ourselves through the measurement of distances and proper motions of hundreds of thousand celestial objects in the nearby universe; where we are and where we will go.  We have to note that the the clocks and scales themselves in the universe, which are indispensable for astrometry, are referenced to distant quasars but affected by a variety of general relativistic effects. We will expect further development of the present-day radio astrometry and astrophysics for the nearby universe in the next decades at the \uas-level accuracy. However, we are simultaneously meeting new challenging issues regarding the time-space reference frame as mentioned above. The dramatic increase in the number of target sources for radio astrometry is necessary to open a new window through those difficulties. This is the basis of our motivation to consider the future global and near-field astrometry with the SKA.     

%\begin{ack}
%Acknowledgement should be placed at end of main text. (NOT after the Appendix.)
%\end{ack}

\bibliographystyle{aa}
\bibliography{paper}

\end{document}